# Stable vortex solitons sustained by a localized gain in the cubic medium

Chunyan Li,[1,2,*] Yaroslav V. Kartashov[2]

[1]School of Physics, Xidian University, Xi'an, 710071, China
[2]Institute of Spectroscopy, Russian Academy of Sciences, 108840, Troitsk, Moscow, Russia

We propose a simple dissipative system with purely cubic defocusing nonlinearity and nonuniform linear gain that can support stable localized dissipative vortex solitons with high topological charges without the utilization of competing nonlinearities and nonlinear gain or losses. Localization of such solitons is achieved due to an intriguing mechanism when defocusing nonlinearity stimulates energy flow from the ring-like region with linear gain to the periphery of the medium where energy is absorbed due to linear background losses. Vortex solitons bifurcate from linear gain-guided vortical modes with eigenvalues depending on topological charges that become purely real only at specific gain amplitudes. Increasing gain amplitude leads to transverse expansion of vortex solitons, but simultaneously it usually also leads to stability enhancement. Increasing background losses allows creation of stable vortex solitons with high topological charges that are usually prone to instabilities in conservative and dissipative systems. Propagation of the perturbed unstable vortex solitons in this system reveals unusual dynamical regimes, when instead of decay or breakup, the initial state transforms into stable vortex soliton with lower or sometimes even with higher topological charge. Our results suggest an efficient mechanism for the formation of nonlinear excited vortex-carrying states with suppressed destructive azimuthal modulational instabilities in a simple setting relevant to a wide class of systems, including polaritonic systems, structured microcavities, and lasers.

**PhySH Subject Headings**: Vortex solitons; dissipative systems

Vortices are ubiquitous topological objects showing intriguing evolution and rich interactions when they are nested in nonlinear fields [1]. Vortices were observed in Bose-Einstein condensates [2,3], hybrid light-matter systems [4], polariton condensates [5-11], in laser systems [12-15], plasmas, and in different optical materials [16-20]. Such states are interesting for practical applications ranging from information encoding, particle trapping, to controllable angular momentum transfer from light to matter. In nonlinear optical media one may observe the formation of vortex solitons. Vortex solitons are excited higher-order nonlinear states [21] that are usually prone to instabilities that may lead to their collapse, decay or splitting into sets of fundamental solitons [22]. Among the strategies allowing to generate stable vortex solitons in conservative optical media is the utilization of materials with competing [23-27] or nonlocal nonlinearities [28-31], various optical potentials [32-40] and other approaches [16,18,22].

Vortex solitons can form not only in conservative, but also in dissipative optical systems, in which case the search of potential stabilization mechanisms becomes particularly important and challenging, since fundamental solitons in such systems are typically characterized by wider attractor basins leading to their predominant dynamical excitation. Nevertheless, dissipative vortex solitons may form in lasers with saturable absorption [12,41-43], in systems governed by complex Ginzburg-Landau equation [44-46], not only in two- [47-52], but also in three-dimensional settings [53-56], in mode-locked lasers [57], and in systems with localized gain and nonlinear absorption [58-66], see reviews [67,68] and recent experimental realizations [69]. In all dissipative systems mentioned above the presence of competing nonlinearities, nonlinear gain or absorption are central for suppression of instabilities of vortex solitons.

In this Letter we propose a new simple mechanism of formation of the ring-like dissipative vortex solitons that does not require competing nonlinearities, nonlinear absorption or optical potentials. Instead, it employs ring-like gain landscape created in a medium with uniform background linear losses and defocusing cubic nonlinearity that in this case prevents an uncontrollable growth of light intensity. Our solitons have localized ring-like shapes despite defocusing nonlinearity and absence of any potentials. They have negative propagation constants laying in the continuous part of the spectrum, where usually only delocalized linear waves exist, thereby illustrating principally different mechanism of formation from that of bright solitons. They bifurcate from gain-guided linear vortex modes existing only for specific set of gain amplitudes, thereby allowing observation of spatial localization over broad range of gain amplitudes due to defocusing nonlinearity. Because in our system stabilization of vortex solitons occurs with increase of gain amplitude, they appear as remarkably robust states that can be stable even for high topological charges, in contrast to vortices in previously considered conservative and dissipative systems, where high-charge solutions are typically unstable.

We consider the propagation of light beams along the $z$-axis in a bulk medium with defocusing cubic nonlinearity and linear ring-like gain landscape $\mathcal{I}(x,y)$, in the presence of background linear losses characterized by the parameter $\alpha$:

$$i\frac{\partial \psi}{\partial z} = -\frac{1}{2}\left(\frac{\partial^2 \psi}{\partial x^2} + \frac{\partial^2 \psi}{\partial y^2}\right) - i[\alpha - \mathcal{I}(x,y)]\psi + |\psi|^2\psi, \qquad (1)$$

where the coordinates $x,y$ are normalized to the characteristic scale $r_0$, the propagation distance $z$ is normalized to the diffraction length $kr_0^2$, $k=2\pi n_\mathrm{r}/\lambda$, intensity is normalized such that $I = n_\mathrm{r}|\psi|^2/k^2r_0^2|n_2|$, where $n_2$ is the nonlinear coefficient. The ring-shaped gain landscape is described by the function $\mathcal{I}(x,y) = \nu e^{-(|\mathbf{r}|-r_\mathrm{c})^2/d^2}$, where $\nu$ is the gain amplitude, $r_\mathrm{c}$ is the radius of the amplifying ring, $d$ is its width, and $\mathbf{r}=(x,y)$. Spatial localization of gain ensures stability of the background at $r\to\infty$. The amplitude of gain/losses $|\alpha-\nu|\sim k^2r_0^2n_\mathrm{i}/n_\mathrm{r}$ is determined by small imaginary part $n_\mathrm{i}$ of the refractive index $n=n_\mathrm{r}+in_\mathrm{i}$, where $n_\mathrm{i}\ll n_\mathrm{r}$. Defocusing cubic nonlinearity is representative for semiconductors, such as $\mathrm{AlGaAs}$ or $\mathrm{CdS}$, for photon energies above $0.7E_\mathrm{bandgap}$ [70-73]. Thus for $\mathrm{CdS}$ at $\lambda=0.61~\mu\mathrm{m}$, where $n_2\approx -10^{-17}~\mathrm{m}^2/\mathrm{W}$ and $n_\mathrm{r}\approx 2.5$, and for $r_0=10~\mu\mathrm{m}$ one gets diffraction length $2.58~\mathrm{mm}$, the amplitude of gain/losses $\alpha,\nu\sim 0.4$ corresponds to $1.55~\mathrm{cm}^{-1}$ (consistent with reported absorption coefficients) while $|\psi|^2=1$ corresponds to $I\sim 3.8\times 10^{12}~\mathrm{W/m}^2$. Different approaches to control of gain in semiconductors have been suggested, based e.g. on electrical or optical pumping or creation of inhomogeneous concentrations of dopants [74]. Even though in this spec-tral range nonlinear absorption also comes into play, it should not affect our solitons at low powers. Effective defocusing cubic nonlinearity can also be produced by cascaded quadratic processes [75]. Models similar to Eq. (1) also supporting vortex solitons may arise in microcavity systems with ring-like pump (see [69,76] and Supplementary Material [77], which includes Refs. [78-81]). Further we set $r_c = 5.25$ and $d = 1.75$. Radially-symmetric vortex soliton solutions of Eq. (1) can be obtained in the form $\psi = (w_r + iw_i)e^{ibz + im\phi}$, where $b$ is the real-valued propagation constant, $m$ is the topological charge, $w_{r,i}(r)$ are the real and imaginary parts of the field. The functions $w_{r,i}$ satisfy the system:

$$bw_{r,i} = \frac{1}{2}\left(\frac{\partial^2}{\partial r^2} + \frac{1}{r}\frac{\partial}{\partial r} - \frac{m^2}{r^2}\right)w_{r,i} + (|w_r|^2 + |w_i|^2)w_{r,i} \mp [\alpha - \mathcal{I}(r)]w_{i,r}. \quad (2)$$

In this dissipative system the propagation constant $b$ is not an independent parameter and it depends on gain/loss amplitude $\alpha, \nu$. To find the profiles $w_{r,i}$ and corresponding $b$ values we used Newton method complemented with an energy flow balance condition that should hold for stationary states:

$$\frac{dU}{dz} = -2\pi \int [\alpha - \mathcal{I}(r)]|\psi|^2 \, rdr = 0, \quad (3)$$

where $U = 2\pi \int |\psi|^2 \, rdr$ is the energy flow of the vortex soliton. Constraint (3) produces additional to (2) equation that is needed in Newton method for definition of propagation constant $b$ (see [77]). Solitons in this system are possible at $\nu > \alpha$, when gain inside the ring becomes sufficiently strong to overcome background losses. Increase of the field amplitude due to amplification within the ring leads to growing defocusing nonlinearity that expels light from the amplifying ring into domain with losses, where energy is absorbed. As a result, a stable energy balance is possible even without nonlinear absorption, allowing to obtain a rich variety of vortex solitons. Due to mechanism of their formation, they are characterized by radial energy currents (besides azimuthal ones associated with vortical phase structure). It should be stressed that vortex solitons are stable only in ring-shaped gain landscapes while in bell-shaped landscapes they are unstable and usually transform into fundamental solitons.

Typical dependencies of the energy flow $U$ on gain amplitude $\nu$ for vortex solitons with different charges $m$ are presented in Fig. 1 at various values of the background losses $\alpha$. Solid lines correspond to stable solitons while dashed lines correspond to unstable ones. Vortex solitons emerge when gain amplitude $\nu$ exceeds certain minimal value depending on topological charge $m$. While at small losses $\alpha = 0.1$ vortices with larger topological charges require larger gain levels for their appearance [Fig. 1(a)], the order of appearance of vortex solitons may change with increase of $\alpha$, so that $m = 2$ [Fig. 1(b)] or $m = 3$ [Fig. 1(c)] vortices may acquire lowest thresholds in $\nu$. This order is determined by the overlap of the field of the vortex with charge $m$ with gain landscape that determines its amplification efficiency. For a given $m$ the threshold value of $\nu$ for appearance of soliton increases with increase of losses $\alpha$. Remarkably, with decrease of $\nu$ the energy flow of vortex soliton vanishes exactly in the point where imaginary part $\lambda_i$ of the complex eigenvalue $\lambda = \lambda_r + i\lambda_i$ of linear eigenmode $\psi = (w_r + iw_i)e^{i\lambda z + im\phi}$ with topological charge $m$ supported by the gain/loss landscape $i[\alpha - \mathcal{I}(r)]$ becomes zero. Such modes are obtained from Eq. (2) with omitted nonlinear term that in this case transforms into linear eigenproblem. Thus, exactly at this value of $\nu$ corresponding linear vortex mode of $i[\alpha - \mathcal{I}(r)]$ landscape evolves without net gain/attenuation. This means that vortex solitons bifurcate from linear gain-guided vortex modes. Notice that while the concept of gain-guiding is well established [82], it was not applied to vortex states. The properties of gain-guided linear vortex modes are summarized in Fig. S1 of [77].

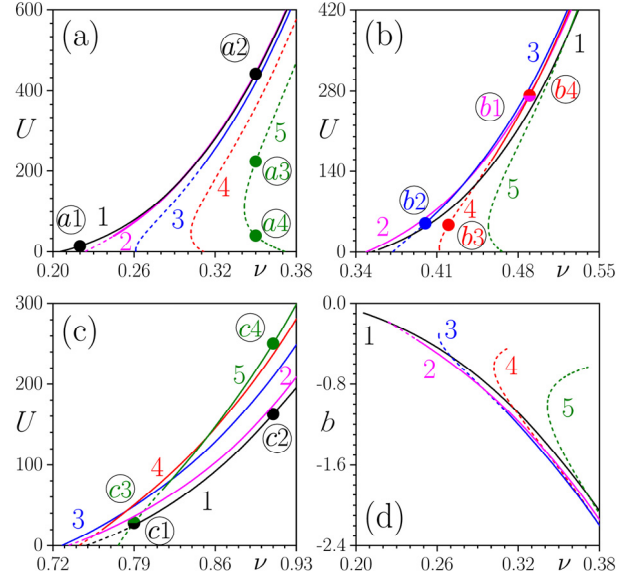

Fig. 1. Energy flow $U$ vs gain amplitude $\nu$ at $\alpha = 0.1$ (a), $\alpha = 0.2$ (b) and $\alpha = 0.5$ (c) for vortex solitons with charges $m = 1, 2, ...5$ indicated next to the curves. (d) Propagation constant $b$ versus $\nu$ at $\alpha = 0.1$. Stable (unstable) branches are shown with solid (dashed) lines. Colored circles and encircled labels correspond to solitons shown in Fig. 2.

Representative dependencies of propagation constants $b$ of vortex solitons with various topological charges on gain amplitude $\nu$ are shown in Fig. 1(d). These dependencies start in the points where soliton's propagation constant $b$ is equal to the eigenvalue $\lambda = \lambda_r$ of the associated gain-guided mode (that is purely real for this $\nu$). Propagation constants of solitons are negative due to defocusing nonlinearity. The dependence $U(\nu)$ may be two-valued, i.e. two different solutions can coexist for the same $\nu$. This usually happens for high topological charges [see $m = 4, 5$ branches in Fig. 1(a)]. With increase of $\alpha$ the dependencies $U(\nu)$ become single-valued for higher and higher $m$ [Fig. 1(c)].

Representative profiles of dissipative vortex solitons corresponding to the dots near encircled labels in Fig. 1(a-c) are presented in Fig. 2 in panels with the same labels. At fixed $\alpha$ with increase of $\nu$ vortex solitons expand in the radial direction far beyond amplifying ring [compare solitons with $m = 1$ in Fig. 2(a1,a2) at $\alpha = 0.1$ or $m = 5$ solitons in Fig. 2(c3),(c4) at $\alpha = 0.5$]. Soliton's width is minimal in the bifurcation point from gain-guided mode. This expansion is accompanied by increase of soliton's amplitude. For high background losses $\alpha \sim 0.5$ the solitons with low charges $m$ may have unusual field modulus distributions $u = (w_r^2 + w_i^2)^{1/2}$ with local radial minima [Fig. 2(c1,c2)]. The presence of radial currents outwards amplifying ring is obvious from oscillations of the tails of the real and imaginary parts $w_{r,i}$ at $r \to \infty$. In the parameter range where the dependence $U(\nu)$ is two-valued, the solitons from lower and upper branches at fixed $\nu$ differ in peak amplitudes and in radial oscillation frequencies of tails of $w_{r,i}$ while widths of the field modulus distributions $u$ may be close [Fig. 2(a3,a4)]. Increasing topological charge $m$ at fixed $\alpha, \nu$ results in increase of the radius of the vortex-ring [compare states in Fig. 2(b1,b4) or states in Fig. 2(c2,c4)]. Higher-order vortex solitons with radial nodes where field modulus $u$ vanishes, were found too (not shown), but all such states are unstable. Qualitatively similar results were obtained for other ring-like $\mathcal{I}(r)$ profiles, for example with step-like gain variation (see Fig. S2 in [77]).

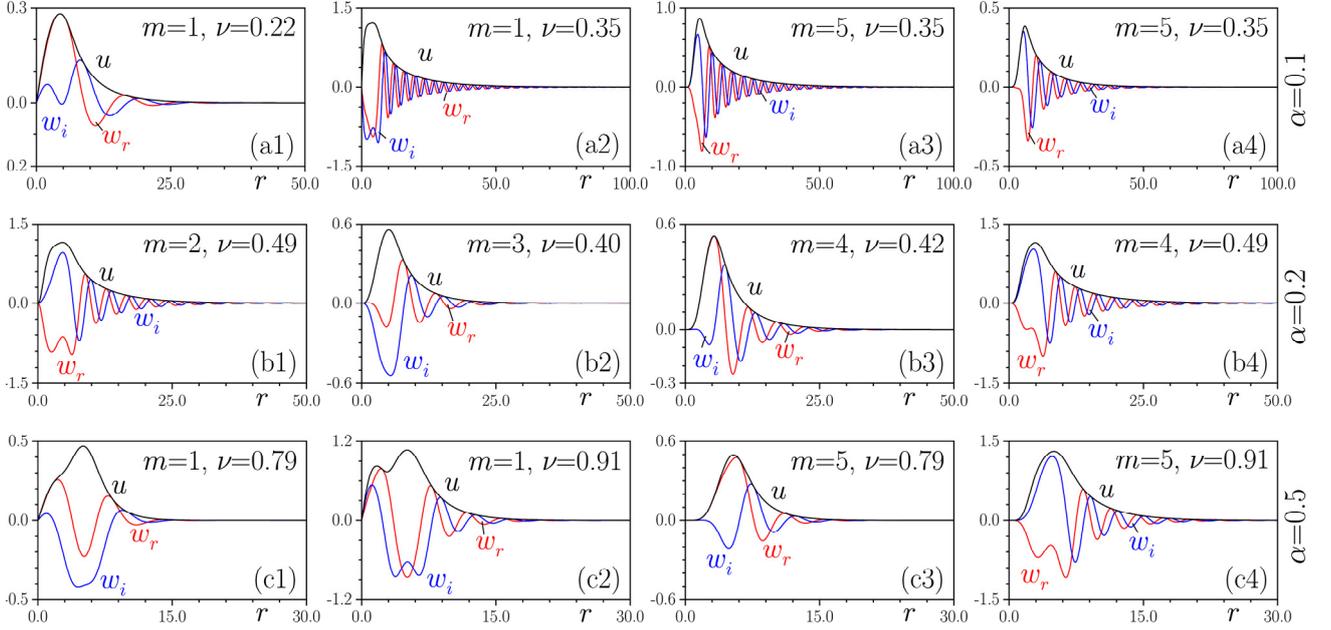

Fig. 2. Vortex solitons with $m=1$ (a1),(a2) and $m=5$ (a3),(a4) at $\alpha=0.1$; $m=2$ (b1), $m=3$ (b2), and $m=4$ (b3),(b4) at $\alpha=0.2$; $m=1$ (c1),(c2) and $m=5$ (c3),(c4) at $\alpha=0.5$, corresponding to labels in Fig. 1(a-c).

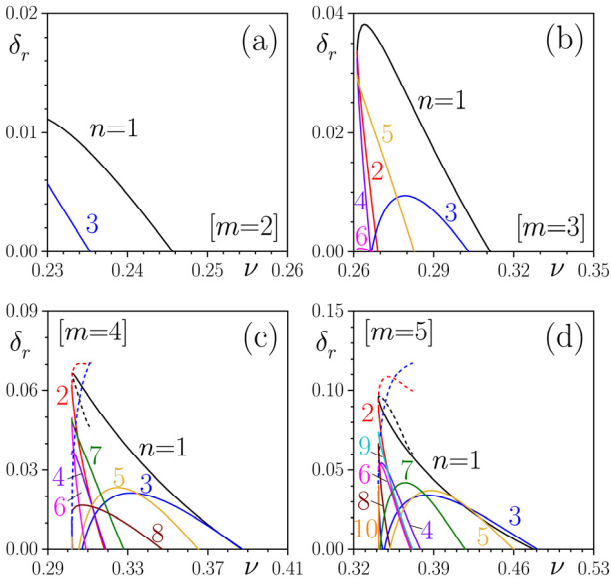

Fig. 3. Real part $\delta_r$ of perturbation growth rate for different azimuthal perturbation indices $n$ versus gain amplitude $\nu$ at $\alpha=0.1$ for solitons with $m=2$ (a), 3 (b), 4 (c), and 5 (d). Dashed curves in (c),(d) correspond to lower branches of the two-valued $U(\nu)$ curves.

The central result of this Letter is that in this simple system, where collapse is absent and azimuthal modulational instabilities are suppressed because material is defocusing, dissipative vortex solitons can be stable even for large topological charges. To illustrate this, we have performed linear stability analysis searching for perturbed states of the form $\psi(r,\phi,z) = (w_r + iw_i + fe^{\delta z + in\phi} + g^* e^{\delta^* z - in\phi})e^{ibz + im\phi}$, where $\delta = \delta_r + i\delta_i$ is the perturbation growth rate, $f(r)$ and $g(r)$ are the radial profiles of perturbation modes, $n$ is the azimuthal perturbation index, and asterisks stand for complex conjugation. The substitution of this ansatz in Eq. (1) and its linearization around $w_r + iw_i$ yields the linear eigenvalue problem for $\delta$ (see [77]), that was solved numerically. Vortex soliton with a given $m$ is stable if $\delta_r \leq 0$ for all $n$ and is unstable otherwise.

The results of stability analysis are summarized in Fig. 1 with vortex soliton families $U(\nu)$ where stable (unstable) families within depicted range of $\nu$ values are plotted with solid (dashed) curves. Representative dependencies of the real part of perturbation growth rate on gain amplitude $\nu$ for different topological charges $m$ or different losses $\alpha$ are shown in Fig. 3 and 4, respectively. We show $\delta_r(\nu)$ dependencies only for azimuthal indices $n$ that can lead to instability within depicted range of $\nu$ values. When instability is present, it usually occurs for low-amplitude solitons near bifurcation point from gain-guided modes (when the dependence $U(\nu)$ is two-valued, its lower branch is usually always unstable, see dashed curves in Fig. 3 and Fig. 4 corresponding to such branches). One of the unusual properties of this system is that at fixed $\alpha$ vortex solitons typically become stable with increase of gain amplitude $\nu$, since growth rates for all $n$ tend to vanish after certain maximal value of $\nu$. Higher-charge vortex solitons usually require larger gain amplitudes $\nu$ for stabilization (see Fig. 3 that shows that the width of the instability domain broadens with increase of $m$), but this picture may change if vortex states with higher charge bifurcate at smaller values of $\nu$ in comparison with $m=1$ states, in which case the former families can be com-pletely stable (Fig. 1). Increasing background losses result in suppression of instabilities associated with large azimuthal perturbation indices $n$ (Fig. 4) leading to narrowing of the instability domain for solitons with sufficiently high topological charges. Thus, at $\alpha=0.9$ even $m=10$ soliton can be stable close to bifurcation point from linear gain-guided state.

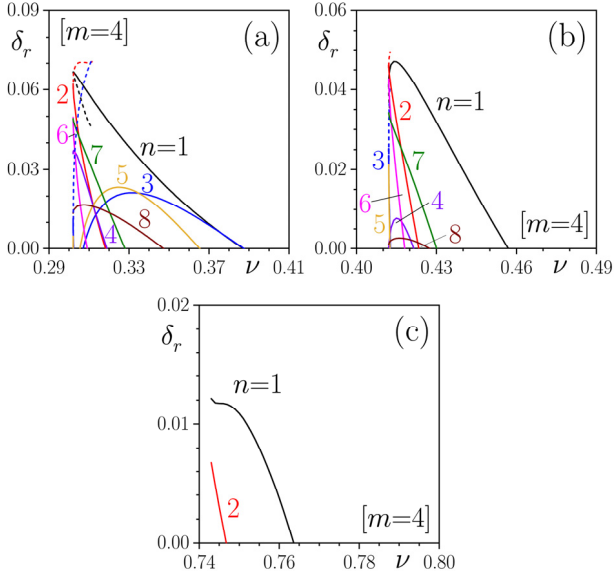

Fig. 4. $\delta_r$ vs $\nu$ dependencies for different azimuthal indices $n$ and vortex soliton with $m=4$ at different values of the background losses $\alpha=0.1$ (a), $\alpha=0.2$ (b), and $\alpha=0.5$ (c).

The existence of stable dissipative vortex solitons was confirmed by direct simulations of their evolution dynamics in the frames of Eq. (1). While stable perturbed vortex solitons do not show any appreciable distortions over huge propagation distances and their peak amplitude $\max|\psi|$ (defined over the entire transverse plane) remains nearly constant indicating on the absence of unstable perturbations [see dynamics of perturbed stable $m=4$ soliton in Fig. 5(c)], the unstable solitons instead of decaying may transform into other stable vortex states not only with lower, but also with higher topological charges. This usually happens when higher-charge states have lower threshold in $\nu$ and are stable while lower-charge states appearing at larger $\nu$ values are unstable [see $m=3$ and $m=1$ branches in Fig. 1(c)]. The example of transformation of unstable $m=1$ vortex into stable $m=3$ state is shown in Fig. 5(a). Another scenario of instability development accompanied by the reduction of topological charge from $m=4$ to $m=2$ is illustrated in Fig. 5(b). In all cases, the vortex solitons emerging as a result of instability development are stable.

In conclusion, we proposed a simple model allowing to obtain remarkably robust dissipative vortex solitons with high topological charges in structured gain/loss landscape combined with defocusing cubic nonlinearity. Being stable attractors of the system such vortex-carrying states, can be easily excited from noisy inputs [77]. Our results pave the way to experimental realization of stable higher-charge vortex solitons in dissipative physical systems, including polaritonic ones, nonlinear structured microcavities, and various laser systems [77]. Vortices are important for practical applications connected with communications, where they can serve as carriers of information encoded in the magnitude/sign of topological charge [83], digital spiral imaging [84], particle manipulation and high-resolution imaging [85,86], and for development of vortex lasers [87]. Robustness of high-charge vortex solitons illustrated in our system and the possibility of transformation between states with different charges may be beneficial for such applications.

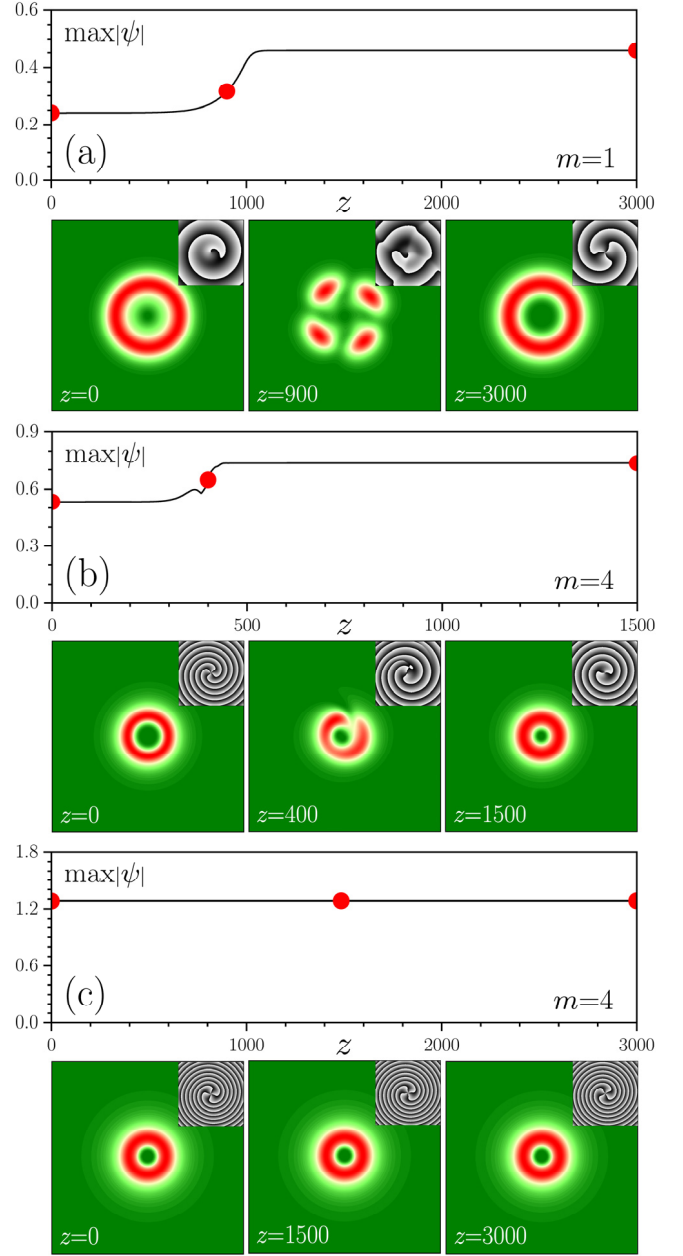

Fig. 5. Evolution dynamics of unstable vortex solitons with (a) $m=1$ at $\alpha=0.5$, $\nu=0.76$ and (b) $m=4$ at $\alpha=0.2$, $\nu=0.42$. Stable evolution of vortex soliton with $m=4$ at $\alpha=0.2$, $\nu=0.51$ (c). Peak amplitude $\max|\psi|$ versus distance $z$ is shown along with field modulus $|\psi|$ and phase $\arg(\psi)$ (insets) distributions at different distances. $(x,y)\in[-15,15]$ in panel (a), $(x,y)\in[-25,25]$ in panels (b) and (c).

**Acknowledgements:** This research is funded by the National Natural Science Foundation of China (NSFC) (11805145); China Scholarship Council (CSC) (202006965016); research project FFUU-2024-0003 of the Institute of Spectroscopy of RAS.

# Supplementary Material: Stable vortex solitons sustained by a localized gain in the cubic medium

Chunyan Li,[1,2,*] Y. V. Kartashov[2]

[1]School of Physics, Xidian University, Xi'an, 710071, China
[2]Institute of Spectroscopy, Russian Academy of Sciences, 108840, Troitsk, Moscow, Russia

**Energy flow balance condition**

Mathematically, the constraint (3) from the main text is obtained by multiplying Eq. (1) with $\psi^*$, subtracting from it complex conjugate equation, and integrating the result in the transverse plane. One can see that the resulting equation

$$\frac{\partial}{\partial z}\iint |\psi|^2 dxdy = -\iint [\alpha - \mathcal{I}(x,y)]|\psi|^2 \, dxdy \qquad \text{(S1)}$$

is the equation describing evolution of beam power $U = \iint |\psi|^2 dxdy$ upon propagation. Since power does not change with distance in stationary states, in polar coordinates and for radially symmetric states and gain, this constraint can be written as

$$2\pi \int [\alpha - \mathcal{I}(r)]|\psi|^2 \, rdr = 0, \qquad \text{(S2)}$$

In Newton method this constraint provides additional equation completing the set of discretized Eq. (2) for soliton profiles and it is necessary for simultaneous iterative calculation of soliton shape $w_r, w_i$ in all $N$ points of the grid and of soliton's propagation constant $b$ that in dissipative systems is not a free parameter and depends on gain amplitude. Therefore in Newton method we store the field $w_r, w_i$ in a vector X with $2N+1$ elements, where first $2N$ elements contain $w_r, w_i$ values on numerical grid while the last element $\mathrm{X}_{2N+1}$ contains unknown $b$ value. Discretized version of the constraint (S2) $\sum_{k=1}^{N}(\alpha - I_k)(w_{r,k}^2 + w_{i,k}^2) = F_{2N+1} = 0$ is then used for the construction of the additional last row $\partial F_{2N+1}/\partial \mathrm{X}_k$ $(k=1,...,2N+1)$ of the Jacobian matrix in Newton method.

**Gain-guided vortex modes in linear system**

Ring-like gain landscape created in the medium with background uniform losses can support localized vortex-carrying states even in the absence of nonlinearity (recall that no usual optical potential is present in this problem and only gain/losses are modulated). These unusual linear gain-guided modes can be found as stationary solutions $\psi = we^{i\lambda z + im\phi}$ of Eq. (1) from the main text without nonlinear term, where $w(r) = w_r + iw_i$ is the complex function describing profile of the mode, $\lambda = \lambda_r + i\lambda_i$ is a complex eigenvalue, and $m$ is the topological charge. Substitution of this field into linear version of Eq. (1) leads to the eigenvalue problem:

$$\lambda w = \frac{1}{2}\left(\frac{\partial^2}{\partial r^2} + \frac{1}{r}\frac{\partial}{\partial r} - \frac{m^2}{r^2}\right)w + i[\alpha - \mathcal{I}(r)]w. \qquad \text{(S3)}$$

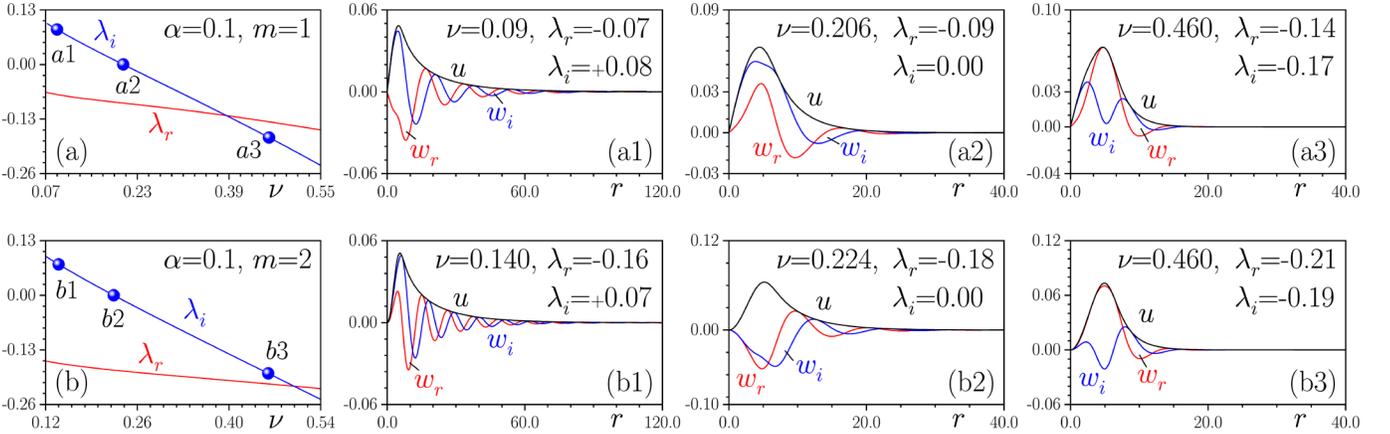

Fig. S1. Real part, $\lambda_r$, and imaginary part, $\lambda_i$, of the eigenvalue of linear gain-guided mode versus gain amplitude $\nu$ for topological charges $m=1$ (a) and $m=2$ (b) at $\alpha=0.1$. Linear mode profiles corresponding to the blue dots in (a) are displayed in (a1-a3) while linear modes corresponding to the blue dots in (b) are displayed in (b1-b3).

The eigenvalue problem (S3) was discretized on numerical grid $[dr/2, 3dr/2, 5dr/2,...]$ to avoid singularity at $r=0$ in two terms $(1/r)\partial/\partial r$ and $m^2/r^2$, and then solved using standard eigenvalue solver. Upon discretization of (S3) in the first point of the grid $k=1$, that corresponds to $dr/2$, the "virtual" element $w_{k-1}$ was replaced either with element $+w_k$ or $-w_k$ in accordance with symmetry of solution $w(r)$ dictated by its topological charge $m$. In linear spectrum of Eq. (S3) one observes the presence of spatially localized gain-guided modes, whose eigenvalues depend on the topological charge $m$, gain amplitude $\nu$, and background losses $\alpha$. Typical dependencies of real $\lambda_r$ and imaginary $\lambda_i$ parts of the eigenvalue of such modes on $\nu$ are depicted in Fig. S1(a) for $m=1$ and Fig. S1(b) for $m=2$ at $\alpha=0.1$. Remarkably, the imaginary part $\lambda_i$ changes from positive (mode is absorbed upon propagation) to negative (mode is amplified upon propagation) values with increase of $\nu$. The particular value of $\nu=0.2059$ for $m=1$ state and $\nu=0.2240$ for

$m=2$ state, where $\lambda_\mathrm{i}$ becomes zero, exactly coincides with the point, where energy flow of vortex soliton with this topological charge vanishes, indicating that vortex solitons bifurcate from gain-guided modes with purely real eigenvalues. Accordingly the propagation constants $b$ of such vortex solitons in bifurcation points [Fig. 1(d)] coincide with real parts of eigenvalues $\lambda_\mathrm{r} = -0.091$ (for $m=1$) and $\lambda_\mathrm{r} = -0.182$ (for $m=2$). Thus, one can conclude that all vortex solitons considered in this Letter bifurcate from gain-guided modes, which, to the best of our knowledge, were never discussed in the frames of gain-guiding concept [1]. Representative profiles of gain-guided modes for different gain amplitudes $\nu$ are presented in Fig. S1(a1-a3) and Fig. S1(b1-b3). These profiles correspond to the blue dots on $\lambda_\mathrm{i}(\nu)$ dependencies in Fig. S1(a) and S1(b), respectively. With decrease of $\nu$ corresponding modes becomes less localized, they develop long tails illustrating the presence of outgoing radial currents from the amplifying ring [see Fig. S1(a1) and (b1)]. Localization progressively increases with the increase of $\nu$, i.e. the modes that experience amplification can be strongly localized [see Fig. S1(a3) and (b3)]. In the bifurcation point for solitons, where $\lambda_\mathrm{i}=0$, gain-guided modes are well-localized too [see Fig. S1(a2) and (b2)] that explains good localization of solitons near bifurcation point.

**Linear stability analysis for vortex solitons**

Linear stability analysis of stationary dissipative vortex solitons is based on the nonlinear Schrödinger equation describing their propagation dynamics [Eq. (1) in the main text], written here in polar coordinates $(r,\phi)$:

$$i\frac{\partial \psi}{\partial z} = -\frac{1}{2}\left(\frac{\partial^2 \psi}{\partial r^2} + \frac{1}{r}\frac{\partial \psi}{\partial r} + \frac{1}{r^2}\frac{\partial^2 \psi}{\partial \phi^2}\right) - i[\alpha - \mathcal{I}(r)]\psi + |\psi|^2\,\psi \quad \text{(S4)}$$

Here we search for perturbed vortex soliton solutions in the form $\psi(r,\phi,z) = (w_\mathrm{r}+iw_\mathrm{i} + fe^{\delta z+in\phi} + g^*e^{\delta^* z - in\phi})e^{ibz+im\phi}$ given also in Eq. (4) in the main text. In this work, we consider all azimuthal perturbation indices $n$ that can potentially lead to instability. Naturally, the range of such $n$ values increases with increase of the topological charge $m$ of the vortex. Substitution of perturbed field in this form into Eq. (S4) and its linearization around stationary solution $w_\mathrm{r} + iw_\mathrm{i}$ yields the linear eigenproblem for complex perturbation growth rate $\delta$:

$$\delta \begin{pmatrix} f \\ g \end{pmatrix} = \begin{pmatrix} i\mathcal{L}_{11} & i\mathcal{L}_{12} \\ -i\mathcal{L}_{21} & -i\mathcal{L}_{22} \end{pmatrix} \begin{pmatrix} f \\ g \end{pmatrix}, \quad \text{(S5)}$$

where the elements of the matrix have the form:

$$\begin{aligned}
\mathcal{L}_{11} &= -b + (1/2)[\partial_r^2 + r^{-1}\partial_r - r^{-2}(m+n)^2] - \\
&\quad 2(w_\mathrm{r}^2 + w_\mathrm{i}^2) + i[\alpha - \mathcal{I}(r)], \\
\mathcal{L}_{12} &= -(w_r + iw_i)^2, \\
\mathcal{L}_{21} &= -(w_r - iw_i)^2, \\
\mathcal{L}_{22} &= -b + (1/2)[\partial_r^2 + r^{-1}\partial_r - r^{-2}(m-n)^2] - \\
&\quad 2(w_\mathrm{r}^2 + w_\mathrm{i}^2) - i[\alpha - \mathcal{I}(r)],
\end{aligned} \quad \text{(S6)}$$

As one can see, the matrix in linear eigenvalue problem (S5) is complex and non-Hermitian, hence, in general, its eigenvalues $\delta$ are complex. Since vortex solitons $(w_\mathrm{r} + iw_\mathrm{i})e^{ibz+im\phi}$ (whose profiles are not available in analytical form) are excited, rather than fundamental nonlinear solutions and $w_\mathrm{r,i} \to 0$ when $r \to 0$, the introduction of analytical stability criterion akin to Vakhitov-Kolokolov criterion for fundamental solitons [2] based on the specific properties of spectrum of linearized operator, is hardly possible in our case. On this reason, linear eigenvalue problem (S5) was solved numerically using standard eigenvalue solver that produces growth rates for different azimuthal perturbation indices. The appearance of perturbations with $\delta_\mathrm{r} > 0$ indicates on the instability of the soliton. Representative dependencies of the perturbation growth rate on gain amplitude $\nu$ are shown in Fig. 3 and 4 in the main text. As one can see from Fig. 1(a)-(c) in the main text, stability properties of vortex solitons may change along the families and this change is not accompanied by any qualitative variations in $U(b)$ or $U(\nu)$ curves. Notice, that due to the presence of background linear losses, the change of stability most frequently occurs when the real part of the eigenvalue $\delta$ crosses zero upon variation of gain amplitude $\nu$, i.e. it does not necessary occur in the point of collision of two purely imaginary eigenvalues, as it frequently happens in conservative systems.

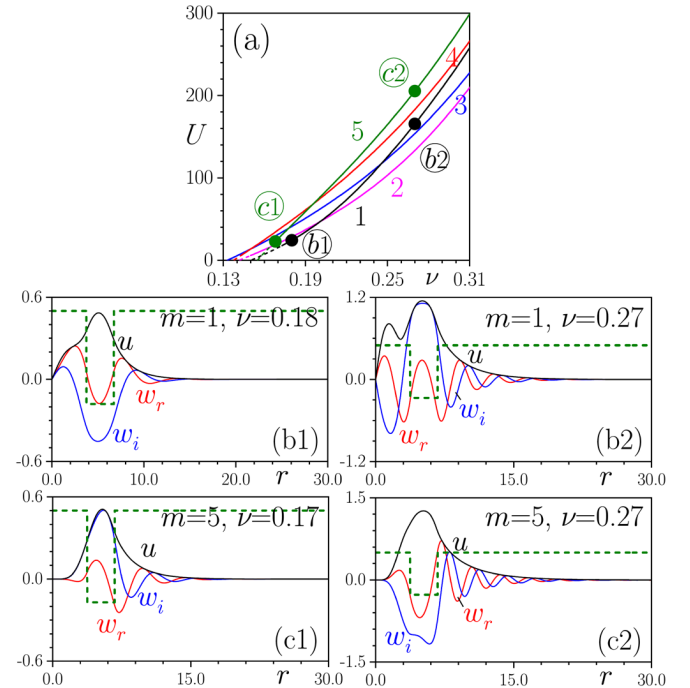

Fig. S2. (a) Energy flow $U$ vs gain amplitude $\nu$ in vortex soliton families with topological charges $m=1,2,\ldots,5$ supported by step-like gain landscape at $\alpha = 0.5$. Topological charges are indicated next to the curves. Stable (unstable) branches in (a) are shown with solid (dashed) lines. Colored circles (and encircled labels next to them) in panel (a) correspond to the profiles of vortex solitons with topological charge $m=1$ depicted in (b1),(b2) and $m=5$ depicted in (c1),(c2). Green dashed lines show $\alpha - \mathcal{I}(r)$ landscape.

We also checked that stable dissipative vortex soliton families can be obtained for other radially-symmetric gain landscapes that stresses generality of the results. In particular, very similar soliton families and stability properties were encountered for ring-shaped gain $\mathcal{I}(r)$ with step-like [i.e. $\mathcal{I}(r)$ is nonzero only at $|r - r_\mathrm{c}| < d/2$ and zero otherwise], rather than Gaussian cross-section. Representative vortex soliton families in such landscapes obtained at $\alpha = 0.5$ are shown in Fig. S2(a), where solid (dashed) segments represent stable (unstable) branches. The examples of soliton profiles corresponding to the dots in Fig. S2(a) are presented in Fig. S2(b1-c2), together with $\alpha - \mathcal{I}(r)$ landscapes depicted with green dashed lines. For these parameters we found only narrow instability domains near the points where energy flow $U$ of corresponding vortex soliton family vanishes. As in the case discussed in the main text, all shown vortex solitons become stable with increase of energy flow. Thus, changing the functional form of $\mathcal{I}(r)$ leads only to quantitative modifications in the existence and stability domains.

**Excitation of vortex solitons from noisy inputs**

Being stable attractors, vortex solitons in our dissipative system can emerge from noisy inputs. The example of such process is illustrated in Fig. S3, where vortex soliton with topological charge $m=3$ gradually builds up from small-scale input noise [see field modulus distributions at different distances $z$ in Fig. S3(b)-(d)] and eventually approaches its final shape that remains unchanged over large distances due to its stability. The evolution of the peak amplitude of the field in this case is illustrated in Fig. S3(a), confirming stability of the emerging state.

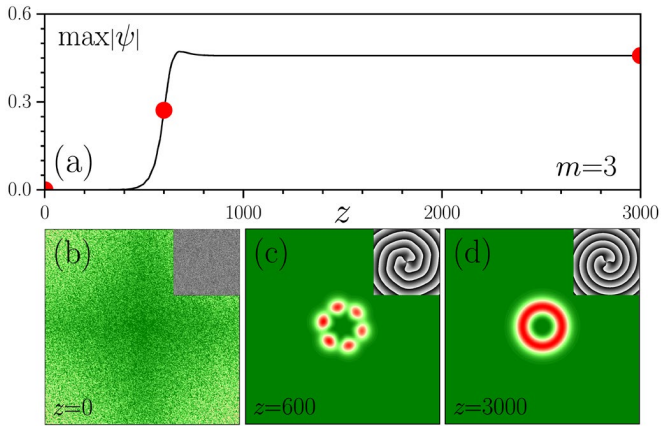

Fig. S3. Spontaneous formation of stable vortex soliton with $m=3$ from small-scale input noise at $\alpha = 0.5$, $\nu = 0.76$. Panel (a) shows peak amplitude $\max|\psi|$ versus distance $z$, while panels (c)-(d) show field modulus $|\psi|$ and phase $\arg(\psi)$ (insets) distributions at different distances within the $(x,y) \in [-25, 25]$ window.

**Connections to model describing polariton condensation in ring-like pump landscapes**

To illustrate that under appropriate conditions the model considered in the main text may be relevant for a broader class of physical systems, as a particular example we consider here planar microcavity operating in strong light-matter coupling regime, where exciton-polaritons can be excited by nonresonant optical pump $P(\mathbf{r})$. In such systems the profile of the pump beam can be easily shaped by using spatial light modulators, see e.g. Refs. [3-5]. The dynamics of polariton condensate is described by the coupled Gross-Pitaevskii equation for the polariton wavefunction $\Psi(\mathbf{r},t)$ coupled to a rate equation for the excitonic reservoir density $n$ that "feeds" the condensate and that is continuously replenished due to the presence of nonresonant optical pump:

$$i\hbar\frac{\partial \Psi}{\partial t} = -\frac{\hbar^2}{2m_{\text{eff}}}\Delta\Psi - i\frac{\hbar}{2}(\gamma_c - Rn)\Psi + g_c|\Psi|^2\Psi + g_r n\Psi,$$
$$\frac{\partial n}{\partial t} = -(\gamma_r + R|\Psi|^2)n + P, \quad \text{(S7)}$$

where $m_{\text{eff}}$ is the effective polariton mass on the lower polariton branch, $\gamma_c$ and $\gamma_r$ are the decay rates of polariton condensate and of the reservoir, parameter $R$ describes stimulated scattering of polaritons from reservoir, and $g_r \approx 2g_c$ is the polariton-reservoir interaction strength. Under the assumption that driven excitonic reservoir adiabatically follows the condensate, one can set $\partial n / \partial t = 0$ and obtain from the last equation that $n \approx P/(\gamma_r + R|\Psi|^2)$. Using Taylor expansion under the assumption $R\gamma_r^{-1}|\Psi|^2 \ll 1$ that is valid not too far from the condensation threshold, one obtains the expression $n = P\gamma_r^{-1}(1 - R\gamma_r^{-1}|\Psi|^2)$. Substituting this expression into Eq. (S7) one obtains the equation for polariton wavefunction $\Psi$:

$$i\hbar\frac{\partial \Psi}{\partial t} = -\frac{\hbar^2}{2m_{\text{eff}}}\Delta\Psi - i\frac{\hbar\gamma_c}{2}\Psi + i\frac{\hbar R}{2\gamma_r}P\Psi + g_c|\Psi|^2\Psi + \frac{g_r}{\gamma_r}P\Psi - i\frac{\hbar R^2}{2\gamma_r^2}P|\Psi|^2\Psi - \frac{g_r R}{\gamma_r^2}P|\Psi|^2\Psi. \quad \text{(S8)}$$

Normalizing spatial coordinates to the characteristic radius $r_0$, one can also introduce characteristic energy $\varepsilon_0 = \hbar^2/m_{\text{eff}}r_0^2$ and time $t_0 = \hbar\varepsilon_0^{-1}$ scales. Then, introducing dimensionless time as $\tau = t/t_0$, the dimensionless wavefunction $\psi = (g_c/\varepsilon_0)^{1/2}\Psi$ and the function describing gain profile $\mathcal{I}(r) = (\hbar R/2\gamma_r\varepsilon_0)R(r)$, one can rewrite Eq. (S8) in dimensionless form:

$$i\frac{\partial \psi}{\partial \tau} = -\frac{1}{2}\left(\frac{\partial^2 \psi}{\partial x^2} + \frac{\partial^2 \psi}{\partial y^2}\right) - i[\alpha - \mathcal{I}(r)]\psi + |\psi|^2\psi + \beta\mathcal{I}(r)\psi - i\mu\mathcal{I}(r)|\psi|^2\psi - \kappa\mathcal{I}(r)|\psi|^2\psi, \quad \text{(S9)}$$

where for experimentally relevant for polariton microcavities parameters $m_{\text{eff}} = 5.63 \times 10^{-5} m_e$, $r_0 = 2~\mu\text{m}$, $\gamma_c = 0.5~\text{ps}^{-1}$, $\gamma_r = 1~\text{ps}^{-1}$, $R = 0.01~\mu\text{m}^2\text{ps}^{-1}$, and $g_r = 2g_c = 4.8~\mu\text{eV}\mu\text{m}^2$, one obtains that the coefficient of linear losses $\alpha = \hbar\gamma_c/2\varepsilon_0 \approx 0.49$, the amplitude of the repulsive potential created by the pump is $\beta = 2g_r/\hbar R \approx 1.46$, the strength of localized nonlinear absorption $\mu = \varepsilon_0 R/\gamma_r g_c \approx 1.41$ and $\kappa = 2g_r\varepsilon_0/\hbar\gamma_r g_c \approx 2.06$. One can see that the terms in the first line of Eq. (S9) are analogous to Eq. (1) in the main text. However, radially-symmetric pump in this system also unavoidably creates local repulsive potential for condensate, induces localized nonlinear losses, and may locally affect nonlinearity in the pumped region. The values of coefficients $\beta, \mu, \kappa$ may vary in the broad range depending on the parameters of microcavity. The Eq. (S9) is valid near the condensation threshold (where amplitude $|\psi|$ start to increase from zero), while sufficiently far from it the condensate density notably increases and one has to resort already to full model (S7) for accurate description of dynamics.

At condensation threshold only linear terms in Eq. (S9) remain. Surprisingly, we have found that gain-guided vortex states appear in this system even in the presence of repulsive potential $\beta\mathcal{I}(r)$ representative for polaritonic system. Such gain-guided linear modes give rise to vortex soliton families depicted in Fig. S4. Here we show only vortex soliton families with topological charges $m=1$ and 3 obtained for the ring-like $\mathcal{I}(r) = \nu e^{-(r-r_c)^2/d^2}$ pump landscape, similar to gain landscape considered in the main text. Fig. S4(a) demonstrates that these solitons are indeed thresholdless, their norm $U$ (defined here analogously to energy flow in the model from main text) monotonically increases with increase of pump amplitude $\nu$. Due to the presence of repulsive potential, such solitons may feature complex density distributions $u^2 = w_r^2 + w_i^2$ with several radial maxima. The amplitude of such solitons vanishes in condensation threshold (different for different topological charges $m$).

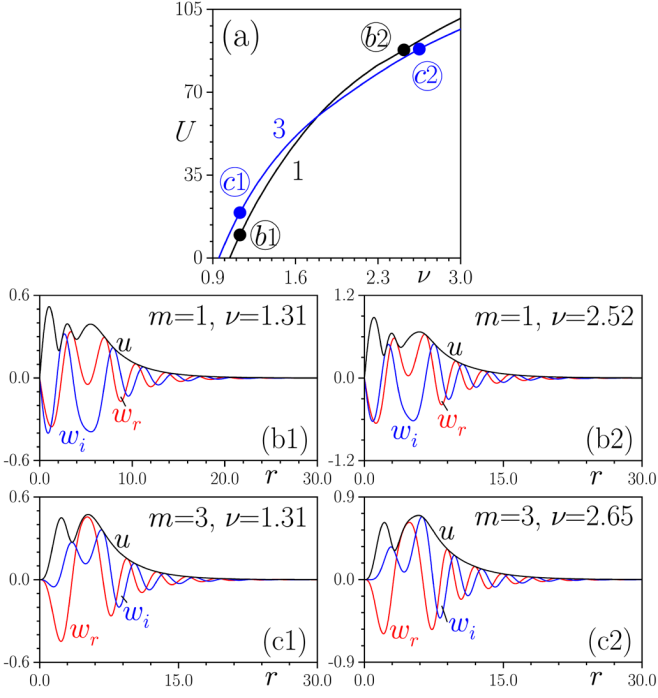

Fig. S4. (a) Norm $U$ vs pump amplitude $\nu$ for vortex solitons with topological charges $m=1,3$ at $\alpha=0.49$ obtained in the frames of the model (S9). Colored dots and labels next to them correspond to vortex solitons with charges $m=1$ (b1),(b2) and $m=3$ (c1),(c2). We show both real $w_r$ and imaginary $w_i$ parts of the wavefunction, as well as its modulus $u = (w_r^2 + w_i^2)^{1/2}$.